\documentclass[twocolumn,showpacs,preprintnumbers,amsmath,amssymb]{revtex4}

\usepackage{lipsum}
\usepackage{tabularx}
\usepackage{array}
\usepackage{mathrsfs}
\usepackage{amssymb}
\usepackage{amsmath}
\usepackage{cases}  
\usepackage{graphicx}
\usepackage{subfigure} 
\usepackage{dcolumn}
\usepackage{bm}
\usepackage{verbatim} 
\usepackage[dvipdfm,pdfstartview=FitH,colorlinks,linkcolor=blue,anchorcolor=blue,citecolor=blue]{hyperref}
\usepackage{amsthm}  

\begin{document}

\title{Non-Markovian property of afterpulsing effect in single-photon avalanche detector}

\author{Fang-Xiang Wang, Wei Chen,\footnote{weich@ustc.edu.cn} Ya-Ping Li, De-Yong He, Chao Wang, Yun-Guang Han, Shuang Wang, Zhen-Qiang Yin and Zheng-Fu Han}
\address{Key Laboratory of Quantum Information, University of Science and Technology of China, Hefei 230026, China\\
and Synergetic Innovation Center of Quantum Information $\&$ Quantum Physics, University of Science and Technology of China,\\
Hefei, Anhui 230026, China\\}


\begin{abstract}

The single-photon avalanche photodiode(SPAD) has been widely used in research on quantum optics. The afterpulsing effect, which is an intrinsic character of SPAD, affects the system performance in most experiments and needs to be carefully handled. For a long time, afterpulsing has been presumed to be determined by the pre-ignition avalanche. We studied the afterpulsing effect of a commercial InGaAs/InP SPAD (The avalanche photodiode model is: Princeton Lightwave PGA-300) and demonstrated that its afterpulsing is non-Markovian, with a memory effect in the avalanching history. Theoretical analysis and experimental results clearly indicate that the embodiment of this memory effect is the afterpulsing probability, which increases as the number of ignition-avalanche pulses increase. This conclusion makes the principle of the afterpulsing effect clearer and is instructive to the manufacturing processes and afterpulsing evaluation of high-count-rate SPADs. It can also be regarded as a fundamental premise to handle the afterpulsing signals in many applications, such as quantum communication and quantum random number generation.

\end{abstract}



\maketitle

\section{Introduction}
\label{Intro.}

Avalanche photodiodes (APDs) are widely used in a variety of fields, such as single molecule detection \cite{Li1993,Moerner2003}, autocorrelated fluorescent decays and quantum information \cite{Eisaman2011,Hadfield2009}, to detect extremely weak light signals or even single photons. The latter is usually named a single-photon avalanche detector (SPAD), which is an essential device and may greatly influence the performance of the system in many applications, especially in quantum information fields \cite{Hadfield2009,Comandar2014}. SPADs usually work in two different modes: free-running mode and gated mode. For free-running mode, the SPAD is biased above its breakdown voltage to allow an avalanche to occur. For gated mode, the SPAD is commonly biased below its breakdown voltage, and the avalanche process can only happen in gating time windows when additional gate voltage signals are superimposed. There are many parameters that can be used to quantify the performance of SPADs, such as the clocking rate, dark count rate, detection efficiency and timing jitter. In the past decade, the performance of SPADs has been greatly improved and they can work at a gigahertz clocking frequency with up to one hundred megahertz counting rates\cite{Yuan2007,Yuan2012,Scarcella2013}.

The afterpulsing effect of SPADs severely limits the count rate and hinders applications that require precise measurements. Afterpulsing, which is correlated with the ignition avalanche, comes from the detrapping of carriers that were trapped at a deep energy level in the junction depletion region \cite{Haitz1965,Cova1991,Cova1996,Kang2003}. The trapped carriers are then released from the traps over time by thermal fluctuation. Usually, the release lifetime $\tau$ is dependent on the type of trap and can vary from 10 $ns$ to several microseconds \cite{Cova1991,Cova2003,Jensen2006}, which is usually much longer than that needed for quenching an avalanche. A releasing carrier may initiate another pseudo detection signal, which is usually called an afterpulse, if the SPAD is biased above the breakdown voltage. The pseudo signal leads to negative effects on SPAD's applications. For example, afterpulsing will increase the quantum bit error rate in the quantum key distribution \cite{Yoshizawa2003,Jain2015} and will destroy the randomness of a quantum random number generator \cite{Wang2015}. To reduce afterpulsing, another hold-off time depending on $\tau$ is necessary. The afterpulsing effect is negligible if the hold-off time is much longer than the carrier lifetime $\tau$, but this is not practical for high-count-rate SPADs. Many approaches have been proposed to reduce afterpulsing \cite{Hadfield2009,Cova1996,Yuan2007,Tosi2014,Wayne2014,Scarcella2015}, but it remains a critical limitation for high-count-rate SPADs. Precisely modeling the behavior of afterpulsing is the necessary foundation to address this problem.

In this paper, we studied the afterpulsing influence of multi-ignition avalanches on the photon detection process of InGaAs/InP SPADs under different gate frequencies and photon intensities and found that afterpulsing possesses a memory effect and thus correlates to the avalanching history. We call this property as non-Markovian property \cite{Bharucha-Reid2012,Markovprocess}. We developed analytical expressions for the afterpulsing influence on photon detection, enabling it to be quantified. We also conducted experiments to verify the theoretical relationships and the experimental results fit well with the theoretical model. The model is instructive to manufacturing processes and the afterpulsing evaluation of high-count-rate SPADs. It also offers a new perspective to address the practical security of quantum communication with imperfect devices \cite{Gottesman2004,Renner2008,Scarani2009}.

\section{Afterpulsing effect analysis}
\label{afterpulsing}

\begin{figure*}[!htb]
\centering
\resizebox{15cm}{5.5cm}{\includegraphics{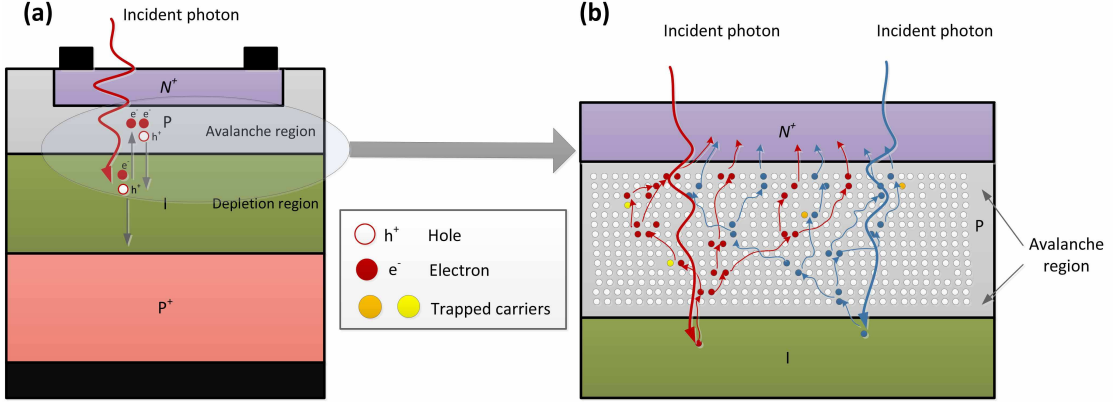}}
\caption{(Color online) (a) Schematic drawing of a SPAD. (b) Enlarge drawing of the elliptic region in (a). It gives a simplified view of the avalanche processes and the increase of carriers (only electrons are shown here). The different colors represent different avalanche processes.}
\label{fig:spad}
\end{figure*}

\subsection{Trapped carriers and afterpulsing}
\label{carriers}
The SPAD (Fig. \ref{fig:spad}(a)) absorbs an incident photon in the depletion region (I region) and creates an electron-hole pair \cite{Eisaman2011}. The electron then drifts into the avalanche region (P region) and triggers an avalanche. The avalanche amplifies the photocurrent so that the initial absorption event can be detected. Some carriers (electrons or holes) are trapped during the avalanche process due to material defects. The release of a carrier in the P region may initiate an afterpulse when the SPAD is biased above the breakdown voltage.

The trap lifetime $\tau$ of carriers in a SPAD can be described by the Arrhenius equation 
\begin{equation}
\tau= C e^{E_{a}/{kT}}
\label{equ1}
\end{equation}
where $E_a$ is the activation energy of the trap, $T$ is the temperature, $k$ is the Boltzmann constant and $C$ is determined by the relevant effective state density, cross-section of the trap and temperature $T$ \cite{Jensen2006}. The carrier release probability is then exponential with respect to the time $t$. As the types of traps in APD are dependent on practical manufacturing processes, there are thereby a variety of traps with different lifetimes $\tau_j$. Thus, the afterpulsing probability (AP) during a detection window $\Delta t$ contains different exponential components \cite{Cova1991}
\begin{equation}
p_a(j)=A_{1}exp(-\frac{j\cdot \Delta t}{\tau_1})+A_{2}exp(-\frac{j\cdot \Delta t}{\tau_2})+\cdots.
\label{equ2}
\end{equation}
where $A_i$ are the amplitudes of different exponential components, and $p_a(j)$ means the AP of the $j$-th detection window after the ignition avalanche. Recently, different fitting methods have been studied that employ a broad distribution of different exponential components instead of summating a few dominant discrete ones \cite{Itzler2012,Horoshko2014}.

So far, most studies on the afterpulsing effect just have focused on its time-dependent probability distribution ignited by a single light pulse and do not discriminate the situation of a single ignition avalanche from that with multiple avalanches \cite{Yoshizawa2002,Kang2003,Yen2008,Silva2011,Stipcevic2013}. However, this is not the real situation when the SPAD works continuously for photon detection. In this case, several avalanches may happen successively within the lifetimes of the trapped carriers. 

Previous works have demonstrated that the number of traps filled with carriers during an avalanche event is much smaller than the total number \cite{Cova1991,Jensen2006,Liu1992}. Thus, avalanches happening successively will increase the number of filled traps. Fig. \ref{fig:spad}(b) gives a simplified view of the filling process. Different photon pulses will be absorbed by different atoms in the I region and the initial electron then drifts into P region. Here, the initial electron randomly collides atoms and each collision may create another unbound electron. The collision makes the paths of the initial and newly unbound electrons unpredictable. These unbound electrons continue the collision process until the edge of the P region. Some defects may be filled by electrons during transfers and collisions. As the filling is rare and the collisions are unpredictable, filling processes of different avalanches affect each other little, as does the release processes. 

\subsection{Non-Markovian property}
\label{non-Markov}

As discussed above, it is reasonable to assume that the filling and release processes of different traps are independent. Thus, avalanches occurring closely will increase the number of trapped carriers and hence will enhance the AP. That is, the AP distribution will be correlated to the avalanching history, while prior studies have not delved down into this physical procedure. In the following paragraphs of this section, we will analyze the AP distribution of an SPAD that avalanches continuously within the lifetime of the dominant carriers.

Considering that the trap level distribution is a technological parameter, which is determined once the product is fabricated and the details have been studied well, we do not consider the specific time-dependent distribution of the AP in the next analysis. The primary data of AP distribution can be obtained by experimental measurements.

The probability of an avalanche triggering by a laser pulse and dark count can be derived from the Poisson distribution, and the avalanche probabilities of different gates (detection windows) have an independent identical distribution (IID) \cite{Wang2015}. Let $p=1-e^{-\eta\lambda}$ be the photon-ignition avalanche probability per gate, $p_a(j)$ be the single-ignition AP of Gate-$j$ and $p_n$ be the total avalanche probability of Gate-$n$, where $\eta$ is the detection efficiency of the SPAD and $\lambda$ is the average photon number per laser pulse. "Single-ignition" means that for a sequence of detection gates, only the first gate (Gate-0) is triggered by the laser pulse. Thus, there is no afterpulsing for Gate-0 and $p_0=p$. 

The AP is determined by the avalanching history. The full-expression of $p_n$, which depends on the avalanching history, is very complex. For example, if Gate-$0$ avalanches, its afterpulsing effect will affect all the following gates. If Gate-$j$ avalanches, its afterpulsing effect will affect the gates after it. The full AP expressions of Gate-$1$ ($p_1$) and Gate-$2$ ($p_2$) are as follows:
\begin{equation}
p_1=p[p+(1-p)p_a(1)]+(1-p)p=p[1+(1-p)p_a(1)],
\label{equ:p1}
\end{equation}

\begin{equation}
\begin{aligned}
p_2=&\quad p[p+(1-p)p_a(1)]\{p+(1-p)[p_a(1)+p_a(2)]\}\\
&+p\{1-[p+(1-p)p_a(1)]\}[p+(1-p)p_a(2)]\\
&+(1-p)p_1.
\end{aligned} 
\label{equ:p2}
\end{equation}
\noindent Here we give a formal expression of $p_n$, and then give the approximate equation of $p_n$. $p_n$ can be classified into two sets: Gate-0 avalanches in the first set $F(p,p_a(j))$ and does not avalanche in the second set $S(p,p_a(j))$. The second set can be expressed as $S(p,p_a(j))=(1-p)p_{n-1}$. $p_n$ can then be expressed as
\begin{equation}
p_n= F(p,p_a(j))+S(p,p_a(j))
\label{equ5}
\end{equation}

If the single-ignition AP $p_a(j)$ is small, and only the first order terms are reserved, namely, $p_a(i)p_a(j)\cong 0$, then we obtain the approximate equation of Equation \ref{equ5},
\begin{equation}
p_n\cong p[1+\sum\limits_{j=1}^{n}(1-p)p_a(j)].
\label{equ6}
\end{equation}
See Appendix for a detailed proof of Equation \ref{equ6}. In fact, the formal expression is enough for the proof.

If the AP is only determined by the latest ignition avalanche \cite{Humer2015}, the afterpulsing is Markovian (has no memory of prior avalanching history). That is, if Gate-$(n-1)$ avalanches, $p(n|n-1)=p+(1-p)p_a(1)$. If Gate-$(n-1)$ does not avalanche, Gate-$(n-2)$ is considered, and so on. Then the approximate equation of $p_n$ becomes
\begin{equation}
p_n\cong p[1+\sum\limits_{j=1}^{n} (1-p)^{j}p_a(j)].
\label{equ4}
\end{equation}
See Appendix for a detailed proof of Equation \ref{equ4}.

Comparing Equation \ref{equ6} with Equation \ref{equ4}, we find that the non-Markovian property makes the afterpulsing effect more significant. 

\section{Experimental setup}
\label{experiment}

\begin{figure}[!tb]
\centering
\resizebox{8.5cm}{6cm}{\includegraphics{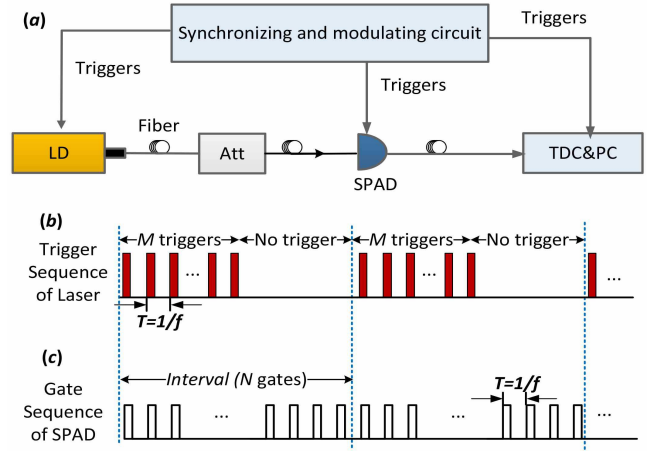}}
\caption{(Color online) Schematic measurement setup of non-Markov property. LD: laser diode; Att: adjustable attenuator; OF: optical fiber; TDC: time-to-digital converter; PC: personal computer.}
\label{fig:setup}
\end{figure}

In this section, we describe experiments to verify the non-Markovian property of afterpulsing.

The experimental setup is shown in Fig. \ref{fig:setup} (a). The laser diode (LD) is operated in pulsed mode. Laser pulses from the LD are attenuated by an adjustable attenuator and are then sent into the SPAD. The SPAD (Princeton Lightwave PGA-300, the operating temperature is $-{50}$ ${}^oC$.) is operated in gated mode without dead time and the gate width is 2.5 $ns$. Its detection efficiency $\eta$ is approximately $10.5\%$. The SPAD outputs a detection signal if an avalanche occurs. The detection signals are input into a time-to-digital converter (TDC, Agilent U1051A Acqiris TC890) in real time. The TDC records the arrival time of every detection signal with a resolution of 50 $ps$. The LD, SPAD and time-to-digital converter (TDC) are triggered and synchronized by a synchronizing and modulating circuit. 

Eq. \ref{equ6} indicates that the non-Markov effect of afterpulsing will be more significant if the single-ignition AP $p_a(j)$ or light intensity increases. To verify Eq. \ref{equ6}, we should measure first the single-ignition AP distribution $p_a(j)$ and second the total avalanche probability $p_n$ of Gate-$n$ with multi-ignition pulses. Fig. \ref{fig:setup}(b) shows the modulating signal of the LD. The number of trigger pulses $M$ (and thus the number of light pulses per interval), trigger frequency $f$ and interval time $\Delta T$ are adjustable. $M=1$ corresponds to the single-ignition case. To measure the AP distribution, the time with no trigger of every interval should be much longer than the carriers' lifetime $\tau$, that is, $t_{interval}-M/f\gg\tau$. The SPAD is triggered continuously by an external pulsed signal (Fig. \ref{fig:setup}(c)) with frequency $f$. Parameter $p$ can be modulated by adjusting the attenuator, and parameter $p_a(j)$ can be modulated by changing the trigger frequency $f$ while keeping the light intensity per pulse constant.

The experiment consists of three scenarios. In \textbf{Scenario (a)}, single-ignition ($M=1$) AP distributions under different trigger frequencies ($f=2, 5, 10$ and $20$ MHz) are measured. The average photon number $\lambda$ per pulse is set as 0.07. The AP $p_a(j)$ is the avalanche probability of Gate-$j$ after subtracting the dark count probability. 

In \textbf{Scenario (b)}, the avalanche probabilities of Gate-$n$ $p_n$ with different average photon numbers per pulse $\lambda$ ($\lambda=1.0, 0.7, 0.07$ and $0.02$) are measured, while the AP distribution $p_a(j)$ varies little. This scenario is used to see how $p_n$ changes with $p$ ($p=1-e^{-\eta\lambda}$). As a contrast, the slope of $p_n$ will decay faster via $p$ if the afterpulsing is Markovian. 

In \textbf{Scenario (c)}, the AP distribution $p_a(j)$ is modulated, while $\lambda$ is kept constant. As discussed in Section \ref{afterpulsing}, the original AP distribution is given by the device technology and operation conditions and does not change. However, according to Equation \ref{equ2}, the value of AP with time, $p_a(j)$, can be changed by re-sampling the detection period $\Delta t$. Hence, in this scenario, the trigger frequency $f$ is modulated ($f=2, 5, 10$ and $20$ MHz) so that the value of $p_a(j)$ becomes different, while $\lambda$ is set as constant ($\lambda=0.07$). This scenario is used to see how $p_n$ changes with $p_a(j)$. This scenario is also used to measure the multi-ignition AP distribution $p_a(j)$, where $p_a(j)$ is the avalanche probability of Gate-$(M+j-1)$. For all scenarios, the dark count rates per gate are under the order of $10^{-5}$. 

\section{results and discussion}
\begin{figure}[!tb]
\centering
\resizebox{8.5cm}{4.5cm}{\includegraphics{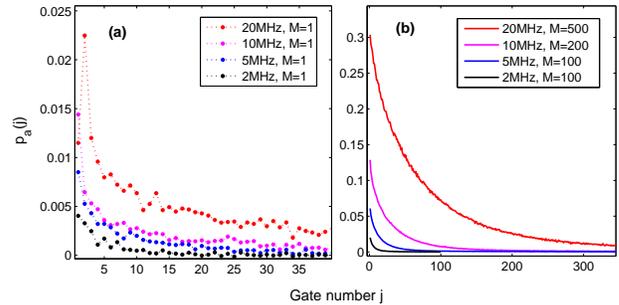}}
\caption{(Color online) Afterpulsing probability distributions with (a) single and (b) multi-ignition pulses. The average photon number per laser pulse is $\lambda=0.07$.}
\label{fig:afpprobdist}
\end{figure}

Single and multi-ignition AP distributions with different triggering frequency $f$ are shown in Fig. \ref{fig:afpprobdist}. The experimental data demonstrate that AP increases as $f$ increases, where $\lambda=0.07$ for all cases. As shown in Fig. \ref{fig:afpprobdist}(a), the total single-ignition AP $p_a=\sum_{j}p_a(j)\approxeq 2.0\%$ when $f=2$ MHz, while $p_a\approxeq 29.9\%$ when $f=20$ MHz. It also demonstrates that multi-ignition APs (Fig. \ref{fig:afpprobdist}(b)) are much larger than the single-ignition ones (Fig. \ref{fig:afpprobdist}(a)). For example, when $f=5$ MHz, the multi-ignition AP of Gate-$(M+j-1)$ is $p_{a,100}(1)=6.1\%$, while the corresponding single-ignition AP is $p_a(1)=0.9\%$. In fact, the total single-ignition AP $p_a\approx\sum\limits_{j=1}^{M=100}p_a(j)=6.2\%\approx p_{a,100}(1)$. This means that the AP accumulates as $M$ increases. Namely, AP depends on the avalanching history and is non-Markovian. The AP becomes larger as $f$ increases. As the full width at half maximum (FWHM) of the detection signal output from the SPAD is approximately 40 $ns$, the saturated processing rate of the discrimination circuit is less than 20 MHz by taking the processing time into account. Thus, $p_a(1)$ is measured to be smaller than the actual value when $f=20$ MHz. However, the processing rate limitation is not that significant for the experiments because the average photon number $\lambda=0.07$, which is small, when $f=20$ MHz.

\begin{figure}[!tb]
\centering
\resizebox{9cm}{6cm}{\includegraphics{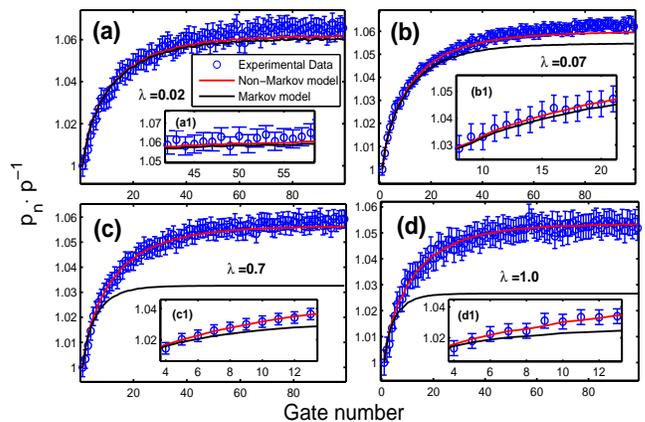}}
\caption{(Color online) Experimental data and theoretical simulations for $p_n$ with different average photon numbers. The trigger frequency is $f=5$ MHz. The insets are partial enlargements. The error bar is set as $\pm 3\sigma$, where $\sigma$ is the standard deviation of the statistics.}
\label{fig:different pa}
\end{figure}

Scenario (b) (see Fig. \ref{fig:different pa}) shows that experimental data (the blue circles) fit Eq. \ref{equ6} (the red lines) well. It proves that the approximate equation is sufficient to describe the non-Markovian property of afterpulsing. As predicted by Eq. \ref{equ4}, the slope of $p_n$ (the black lines) of the Markovian model decays faster than that of the experimental data. The difference between the non-Markovian model and Markovian model is negligible when the photon-ignition avalanche probability $p$ ($\lambda=0.02, p=1-e^{-\eta\lambda}\approx 0.002$) is small enough. The difference becomes more significant when $p$ increases, e. g., Fig. \ref{fig:different pa}(d) shows that $p_{{}_{n=99}}^{non-Markov}-p_{{}_{n=99}}^{Markov}=1.053\cdot p-1.027\cdot p=0.026\times p$. The Markovian model cannot fit with the experimental data. 

\begin{figure}[!tb]
\centering
\resizebox{9cm}{6cm}{\includegraphics{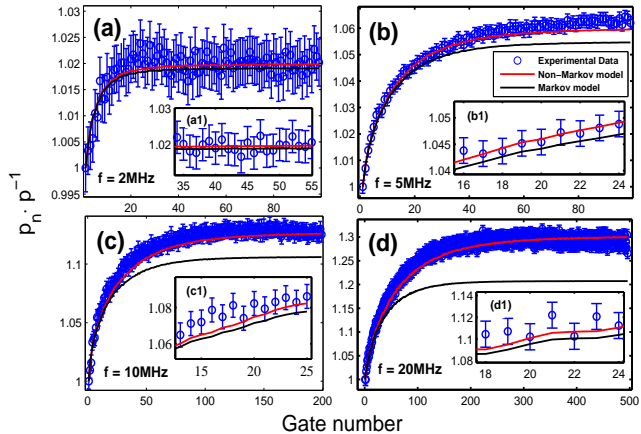}}
\caption{(Color online) Experimental data and theoretical simulations for $p_n$ with different gate frequency $f$. The average photon number per laser pulse is $\lambda=0.07$. The detection efficiencies are $\eta_{{}_{f=2MHz}}=10.3\%,\eta_{{}_{f=5MHz}}=10.5\%,\eta_{{}_{f=10MHz}}=11.6\%,\eta_{{}_{f=20MHz}}=12.0\%$. The insets are partial enlargements. The error bar is also set as $\pm 3\sigma$.}
\label{fig:different gate}
\end{figure}

Scenario (c) provides differences of $p_n$ between the non-Markovian and Markovian models with different $p_a(j)$ (see Fig. \ref{fig:different gate}). There is hardly any difference between these two model when $f=2$ MHz ($p_a=2.0\%$), as shown in Fig. \ref{fig:different gate}(a). However, the difference becomes $p_{n}^{non-Markov}-p_{n}^{Markov}\approx 0.1\cdot p$ for large $n$ when $f=20$ MHz ($p_a=29.9\%$).

The experimental results above are all well consistent with the non-Markovian analysis. According to Eq. \ref{equ6}, the afterpulsing influence on photon detection depends on the product of the photon-ignition avalanche probability and the total AP. In both Fig. \ref{fig:different pa}(a) and Fig. \ref{fig:different gate}(a), $p\cdot p_a$ is very small (approximately $10^{-4}$), and the Markov model also fits well with the experimental data, although the assumption is incorrect. However, in Fig. \ref{fig:different pa}(b),(c),(d) and \ref{fig:different gate}(b),(c),(d), the Markovian model can no longer fit with the data, as $p\cdot p_a$ becomes larger (about $10^{-3}$ in \ref{fig:different pa}(d) and \ref{fig:different gate}(d)). The non-Markovian effect of afterpulsing will be more significant for higher $p_a$. This is really the case for high-count-rate SPADs. 

It should be noted that, due to the limit of the designed processing circuit, the detection efficiency of the SPAD varies with the trigger frequency $f$. To make a comparison of the AP influence with different frequencies, the gate voltage loaded on the SPAD is fine adjusted so that detection efficiency varies less ($\eta_{{}_{f=2MHz}}=10.3\%,\eta_{{}_{f=5MHz}}=10.5\%,\eta_{{}_{f=10MHz}}=11.6\%,\eta_{{}_{f=20MHz}}=12.0\%$). Otherwise, afterpulsing would be a catastrophe and the SPAD under $f=20$ MHz could no longer work properly if the voltage remains the same as that under $f=2$ MHz. By utilizing high-speed (e.g., 100 MHz) processing circuit with improved design, the detection efficiency will vary less under the same conditions.

Early works have demonstrated that the number of traps filled with carriers during an avalanche event is much smaller than the total number in both visible and near-infrared APD \cite{Cova1991,Jensen2006,Liu1992}. Thus, the afterpulsing effects of these types of SPADs can be described by the non-Markovian model shown in Figure \ref{fig:spad}. Recently, afterpulsing effect is still a serious limit toward high-count-rate SPADs \cite{Tosi2014,Wayne2014}. The SPADs with gigahertz clocking frequency and one hundred megahertz counting rate are mainly reached by circuit improvement, by which the gate voltage, total avalanche current and the AP have been suppressed significantly \cite{Yuan2007,Yuan2012,Scarcella2015}.  However, the material quality has not been improved much in the last decade \cite{Italer2011} and the defects of the materials are still in large quantity \cite{Tosi2014}. Hence, despite that only a particular commercial InGaAs/InP SPAD has been verified in this paper, the non-Markovian model is general for SPADs unless a significant breakthrough of the fabrication technology is being made. Our work reveals that the circuit quenching method, which cannot change the intrinsic non-Markovian property of afterpulsing, will remain one of the bottlenecks for higher-count-rate SPADs. The once and for all solution is to reduce the defects in the materials or to erase the trapped carriers before the arrival of the next photon.

\section{Conclusions}
\label{conclusion}

In conclusion, we have proved that the afterpulsing effect of the SPAD has a memory of the avalanching history and is hence non-Markovian. The approximate expression (Eq. \ref{equ6}) well describes the non-Markovian property of afterpulsing. The non-Markovian effect will be more significant for higher $p_a$. This is the case for high-count-rate SPADs, where the detection period is much smaller than the lifetime of the carriers \cite{Yuan2007,Yuan2012,Scarcella2013}. As the embodiment of rarely filling of traps, the non-Markovian property of afterpulsing should be general to different types of SPADs provided that the filling is rare during an avalanching process. Our work makes the principle of the afterpulsing effect clearer and is instructive to manufacturing processes and the afterpulsing evaluation of high-count-rate SPADs. It also offers a new perspective to consider the practical security of quantum communication due to the imperfection of devices.

%


\setcounter{equation}{0}
\section*{Appendix}
\label{Appendix}

\subsection{The Markovian approximate equation}

If AP distribution is only determined by the latest ignition avalanche as assumed in \cite{Humer2015},namely, the AP distribution has no memory of prior avalanching history and is Markovian. That is, if Gate-$(n-1)$ avalanches, $p(n|n-1)=p+(1-p)p_a(1)$. If Gate-$(n-1)$ does not avalanche, Gate-$(n-2)$ is considered. If Gate-$(n-2)$ avalanches, $p(n|(\overline{n-1},n-2))=p+(1-p)p_a(2)$. If Gate-$(n-2)$ does not avalanche, Gate-$(n-3)$ is considered, and so on. Hence, the avalanche probability of Gate-$n$ is
\begin{equation}
\begin{aligned}
p_n=&\quad p_{n-1}[p+(1-p)p_a(1)]\\
&+(1-p_{n-1})\{p_{n-2}[p+(1-p)p_a(2)]+(1-p_{n-2})\\
&\quad\{\cdots\{p_0[p+(1-p)p_a(n)]+(1-p_0)p\underbrace{\}\cdots\}\}}_{(n-1)"\}"}.
\end{aligned}
\label{equ11}
\end{equation}
The first term of the equation means if the pre-gate avalanches, the ones before the pre-one are not considered. If the pre-gate does not avalanche,  the gate before the pre-gate is under considered, and so on. 

By keeping the first-order terms of $p_a(j)$ only, we obtain the approximate equation
\begin{equation}
p_n\cong p[1+\sum\limits_{j=1}^{n} (1-p)^{j}p_a(j)].
\label{equ12}
\end{equation}

\begin{proof}

First, we consider the limiting case of no afterpulsing. In this case, $p_n=p$ always holds.

Thus, in general, $p_n$ can be expressed as $p_n=p+g_n(p_a(j))$, where $g_n(p_a(j))=g_n(p_a(1),p_a(2),p_a(3),\cdots)$ is a power series of $p_a(j)$ and $j=1,2,\cdots$. For the first term of right side, we have
\begin{equation}
\begin{aligned}
&\quad p_{n-1}[p+(1-p)p_a(1)]\\
&=[p+g_{n-1}(p_a(j))]\cdot[p+(1-p)p_a(1)]\\
&=p^2+p(1-p)p_a(1)+pg_{n-1}(p_a(j))\\
&\quad +g_{n-1}(p_a(j))\cdot (1-p)p_a(1)\\ 
&\cong p^2+p(1-p)p_a(1)+pg_{n-1}(p_a(j)).
\end{aligned}
\label{equ13}
\end{equation}
The remainder terms become
\begin{equation}
\begin{aligned}
&(1-p_{n-1})\{p_{n-2}[p+(1-p)p_a(2)]+(1-p_{n-2})\\
&\{\cdots\{p_0[p+(1-p)p_a(n)]+(1-p_0)p\underbrace{\}\cdots\}\}}_{(n-1)"\}"}\\
=& [1-p-g_{n-1}(p_a(j))]\{p_{n-2}[p+(1-p)p_a(2)]+(1-p_{n-2})\\
&\{\cdots\{p_0[p+(1-p)p_a(n)]+(1-p_0)p\underbrace{\}\cdots\}\}}_{(n-1)"\}"}\\
=& (1-p)\{p_{n-2}[p+(1-p)p_a(2)]+(1-p_{n-2})\\
&\{\cdots\{p_0[p+(1-p)p_a(n)]+(1-p_0)p\underbrace{\}\cdots\}\}}_{(n-1)"\}"}\\
&-g_{n-1}(p_a(j))\{p_{n-2}[p+(1-p)p_a(2)]+(1-p_{n-2})\\
&\{\cdots\{p_0[p+(1-p)p_a(n)]+(1-p_0)p\underbrace{\}\cdots\}\}}_{(n-1)"\}"}.
\end{aligned}
\label{equ14}
\end{equation}
In order to simplify term $g_{n-1}(p_a(j))\{p_{n-2}[p+(1-p)p_a(2)]+(1-p_{n-2})\{\cdots\{p_0[p+(1-p)p_a(n)]+(1-p_0)p\underbrace{\}\cdots\}\}}_{(n-1)"\}"}$, we begin with the innermost brace. The term containing $p_a(n)$ can be abandoned directly by the approximation convention $p_a(i)p_a(j)\cong 0$. Then, the expression of the innermost brace is simplified to $p$. Then, the expression outside the innermost brace becomes $p_1[p+(1-p)p_a(n-1)]+(1-p_1)p\cong p$. The expressions of outer braces can be simplified by analogy, and $g_{n-1}(p_a(j))\{p_{n-2}[p+(1-p)p_a(2)]+(1-p_{n-2})\{\cdots\{p_0[p+(1-p)p_a(n)]+(1-p_0)p\underbrace{\}\cdots\}\}}_{(n-1)"\}"}\cong g_{n-1}(p_a(j))p$. Equation \ref{equ11} hence becomes
\begin{equation}
\begin{aligned}
p_n=
& p^2+p(1-p)p_a(1)+(1-p)\{p_{n-2}[p+(1-p)p_a(2)]\\
&+(1-p_{n-2})\{\cdots\{p_0[p+(1-p)p_a(n)]+(1-p_0)p\\
&\quad\underbrace{\}\cdots\}\}}_{(n-1)"\}"}\\
\end{aligned}
\label{equ15}
\end{equation}
Analogously, $p_{n-2}, p_{n-3}, \cdots$ can be simplified by the same procedure done for $p_{n-1}$. Equation \ref{equ11} finally becomes
\begin{equation}
\begin{aligned}
p_n=& p^2+(1-p)p^2+(1-p)^{2}p^2+\cdots +(1-p)^{n-2}p^2\\
&+(1-p)^{n-1}p+p(1-p)p_a(1)+p(1-p)^{2}p_a(2)\\
&+\cdots+p(1-p)^{n-1}p_a(n-1)+p_0(1-p)^{n}p_a(n)\\
=& p[1+\sum\limits_{j=1}^{n}(1-p)^{j}p_a(j)].
\end{aligned}
\label{equ16}
\end{equation}
This completes the proof.
\end{proof}

There is an alternative proof of the Markovian approximate equation. The proof is shown below.

\begin{proof}
 The recurrence relation of the sequence $p_n$ can be written as follows according to  Equation \ref{equ11}:

\begin{equation}
\begin{aligned}
p_{n+1}=& p+p_{n}(1-p)p_a(1)+p_{n-1}(1-p_{n})(1-p)p_a(2)\\
&+\cdots+p(1-p_1)\cdots (1-p_{n})(1-p)p_a(n+1)
\end{aligned}
\label{equ17}
\end{equation}

According to Equation \ref{equ17}, it is easy to get that $p_1=p +p(1-p)p_a(1)$, $p_2=p+p1(1-p)p_a(1)+p(1-p1)(1-p)p_a(2)\cong p[1+(1-p)p_a(1)+(1-p)^2p_a(2)]$, where we only keep the first-order terms of $p_a(j)$.  Thus, we assume that $p_n\cong p[1+\sum\limits_{j=1}^{n} (1-p)^{j}p_a(j)]$. If $p_{n+1}=p_{n}+p(1-p)^{n+1}p_a(n+1)$ holds, then Equation \ref{equ12} holds.

According to the expression of $p_{n}$, we can replace all the $p_{n}$ in Equation \ref{equ17} by $p$ since we neglect higher than the first-order terms of $p_a(j)$. Thus we can get straightforwardly:

\begin{equation}
\begin{aligned}
p_{n+1}\cong& p+p(1-p)p_a(1)+p(1-p)^2 p_a(2)\\
&+\cdots +p(1-p)^{n+1}p_a(n+1)\\
=& p[1+\sum\limits_{j=1}^{n+1} (1-p)^{j}p_a(j)]
\end{aligned}
\label{equ18}
\end{equation}

This means Equation \ref{equ12} holds for $n+1$, thus completes the proof.

\end{proof}

\subsection{The non-Markovian approximate equation}

The formal expression of avalanche probability of Gate-$n$ in non-Markovian model is
\begin{equation}
p_n= F(p,p_a(j))+S(p,p_a(j))
\label{equ19}
\end{equation}

By keeping the first-order terms of $pa(j)$ only, we obtain the approximate equation of non-Markovian model
\begin{equation}
p_n\cong p[1+\sum\limits_{j=1}^{n}(1-p)(p_a(j))].
\label{equ20}
\end{equation}

\begin{proof}
According to Equation \ref{equ19}, it is easy to get that $p_1=p[1+(1-p)p_a(1)]$, $p_2=p[1+(1-p)(p_a(1)+p_a(2))+(1-p)^2{p_a(1)}^2]\cong p[1+(1-p)(p_a(1)+p_a(2))]$, and $p_3=p[1+(1-p)(p_a(1)+p_a(2)+p_a(3))+(1-p)^2{p_a(1)}^2+2(1-p)^2p_a(1)p_a(2)+(1-p)^3{p_a(1)}^3] \cong p[1+(1-p)(p_a(1)+p_a(2)+p_a(3))]$. Thus, we assume that $p_{n-1}=p[1+(1-p)(p_a(1)+p_a(2))+\cdots+p_a(n-1)]$. If $p_n=p_{n-1}+p(1-p)p_a(n)$ holds, then Equation \ref{equ20} holds.

The right side of Equation \ref{equ19} includes two sets: Gate-0 avalanches in the first set $F(p,p_a(j))$ and does not avalanche in the second set $S(p,p_a(j))$. The second set can be expressed as $S(p,p_a(j))=(1-p)p_{n-1}$. Equation \ref{equ19} can then be expressed as
\begin{equation}
\begin{aligned}
p_n=&\underbrace{p}_{Gate-0}\cdot\underbrace{[p+(1-p)p_a(1)]}_{Gate-1}\cdot\underbrace{\{\cdots\}}_{f_1(p,p_a(j))}\\
&+\underbrace{p}_{Gate-0}\cdot\underbrace{[1-p-(1-p)p_a(1)]}_{Gate-1}\cdot\underbrace{\{\cdots\}}_{f_1^{'}(p,p_a(j))}\\
&+\underbrace{\underbrace{(1-p)}_{Gate-0}\cdot\underbrace{p_{n-1}}_{\ Remaining\ terms}}_{S(p,p_a(j))}.
\end{aligned}
\label{equ21}
\end{equation}
The main difference of $F(p,p_a(j))$ from $S(p,p_a(j))$ is the afterpulsing effect ignited by Gate-0. If we separate out all terms due to the afterpulsing effect of Gate-0 (that is, terms with factors $p_a(1)$ of Gate-1, $p_a(2)$ of Gate-2, $\cdots$, $p_a(n)$ of Gate-$n$), the remaining terms of $F(p,p_a(j))$ can be expressed as $p\cdot p_{n-1}$. According to Equation \ref{equ21}, terms with factor $p_a(1)$ of Gate-1 can be expressed in the form of $p(1-p)p_a(1)f_1(p,p_a(j))-p(1-p)p_a(1){f_1}^{'}(p,p_a(j))$, where $f_1(p,p_a(j))$ and ${f_1}^{'}(p,p_a(j))$ are the abbreviations for the remaining terms. According to the approximate convention, all terms with $p_a(j)$ in  $f_1(p,p_a(j))$ and ${f_1}^{'}(p,p_a(j))$ are abandoned. Then $f_1(p,p_a(j))$ and ${f_1}^{'}(p,p_a(j))$ approximate to the same expression 
\begin{equation}
\begin{aligned}
f_1(p,p_a(j))\cong& {f_1}^{'}(p,p_a(j))\\
\cong& {{n-1}\choose 0}p^{n-1}+{{n-1}\choose 1}p^{n-2}(1-p)\\
&+\cdots+{{n-1}\choose {n-1}}(1-p)^{n-1}\\
=&[p+(1-p)]^{n-1}\\
=& 1,
\end{aligned}
\label{equ22}
\end{equation}
where ${n-1}\choose k$ is the binomial coefficient. Hence, $p(1-p)p_a(1)f_1(p,p_a(j))-p(1-p){f_1}^{'}(p,p_a(j))\cong 0$. Analogously, Terms with factor $p_a(2)$ of Gate-2, $p_a(3)$ of Gate-3, $\cdots$, $p_a(n-1)$ of Gate-$(n-1)$ all approximate to 0. As there is no term $-(1-p)p_a(n)$ for Gate-$n$, there remains a nonzero term $p(1-p)p_a(n)f_n(p,p_a(j))\cong p(1-p)p_a(n)$, where $f_n(p,p_a(j))\cong 1$ by the approximation convention. Thus,
\begin{equation}
\begin{aligned}
p_n&=F(p,p_a(j)+S(p,p_a(j))\\
&=p\cdot p_{n-1}+p(1-p)p_a(n)+(1-p)p_{n-1}\\
&=p_{n-1}+p(1-p)p_a(n)\\
&=p[1+\sum\limits_{j=1}^{n}(1-p)(p_a(j))].
\end{aligned}
\label{equ23}
\end{equation}
This completes the proof.

\end{proof}


\begin{thebibliography}{1}
\bibitem{Markovprocess}A process satisfies the Markovian property if one can make predictions for the future of the process based solely on its present state, i. e., conditional on the present state of the system, its future and past are independent. Otherwise, if its future and past are correlated, the process is non-Markov. As shown in this letter, the afterpulsing probability distribution depends on the avalanching history. Thus afterpulsing effect satisfies non-Markovian property.
\end{thebibliography}
\end{document}